\documentstyle[12pt]{article}
\if@twoside
\else
\fi
\advance\textheight by \topskip
\leftmargin\leftmargini
\labelwidth\leftmargini\advance\labelwidth-\labelsep

\catcode`\@=11
\def\Let@{\relax\iffalse{\fi\let\\=\cr\iffalse}\fi}
\def\vspace@{\def\vspace##1{\crcr\noalign{\vskip##1\relax}}}
\def\multilimits@{\bgroup\vspace@\Let@
 \baselineskip\fontdimen10 \scriptfont\tw@
 \advance\baselineskip\fontdimen12 \scriptfont\tw@
 \lineskip\thr@@\fontdimen8 \scriptfont\thr@@
 \lineskiplimit\lineskip
 \vbox\bgroup\ialign\bgroup\hfil$\m@th\scriptstyle{##}$\hfil\crcr}
\def\Sb{_\multilimits@}
\def\endSb{\crcr\egroup\egroup\egroup}
\def\Sp{^\multilimits@}

\def\stackunder#1#2{\mathrel{\mathop{#2}\limits_{#1}}}
\long
\def\QQQ#1#2{}
\def\QTP#1{}
\long
\def\QQA#1#2{}

\def\EXPAND#1[#2]#3{}
\def\NOEXPAND#1[#2]#3{}

\def\LaTeXparent#1{}

\QQQ{Language}{
American English
}

\begin{document}

$ $

\vskip 3cm

\begin{center}
\LARGE{Generalization of the Hamilton-Jacobi approach for higher order singular
systems} 
\end{center}

\,

\begin{center}
B. M. Pimentel and R. G. Teixeira\\
\end{center}

\,

\begin{center}
Instituto de F\'{\i}sica Te\'orica\\Universidade Estadual Paulista\\Rua
Pamplona 145\\01405-900 - S\~ao Paulo, S.P.\\Brazil\\
\end{center}

\vskip 1cm

\begin{center}
\begin{minipage}{14.5cm}
\centerline{\bf Abstract}
\,
We generalize the Hamilton-Jacobi formulation for higher order singular systems and 
obtain the equations of motion as total differential equations. To do this we first 
study the constrains structure present in such systems.
\end{minipage}
\end{center}

\newpage\ 

\section{Introduction}

The study of singular systems has reached a great status in physics since
the development by Dirac \cite{Dirac 1, Dirac 2, Dirac 3} of the generalized
Hamiltonian formulation. Since then, this formalism has found a wide range
of applications in Field Theory \cite{Regge, Sundermeyer, Gitman, Teitelboim}
and it is still the main tool for the analysis of singular systems. Despite
the success, it is always interesting to apply different formalisms to the
analysis of singular systems since they may show new features of the system
under study, in a similar way to what happens in Classical Dynamics \cite
{Lanczos}.

Recently an approach based on Hamilton-Jacobi formalism was developed to
study singular first order systems \cite{Guler 1, Guler 2}. This approach
consists in using Ca\-ra\-th\'{e}o\-do\-ry's equivalent Lagrangians method
to write down the Hamilton-Jacobi equation for the system and make use of
its singularity to write the equations of motion as total differential
equations in many variables. This new approach, due to its very recent
development, has been applied to very few examples \cite{Guler 3, Guler 4,
Guler 5, Guler 6} and it is still necessary a better understanding of its
features, its advantages and disadvantages in the study of singular systems
when compared to Dirac's Hamiltonian formalism.

Besides that, theories with higher order Lagrangians (or higher order wave
equations) are important in the context of many physical problems. Examples
range from Podolsky's Generalized Electrodynamics \cite{Podolsky} to
tachyons (ref. \cite{Bollini} and references there in).

Our aim here is to make a formal generalization of Hamilton-Jacobi formalism
for singular systems with arbitrarily higher order Lagrangians. This
generalization is motivated by the attention that higher order systems have
received in literature \cite{Nesterenko, Batlle, Saito}. A treatment for the
case of second order Lagrangians has already been developed by the authors 
\cite{Eu}, but here we will make a more general treatment that will begin
with the analysis of the constraints' structure of such systems (Sect. 2).
Next we will use Carath\'{e}odory's equivalent Lagrangians method to derive
the Hamilton-Jacobi equation for higher order systems (Sect. 3) and analyze
the singular case (Sect. 4). The conclusions will be drawn in Sect. 5.

\section{Constraints structure in higher order systems}

We will analyze a system described by a Lagrangian dependent up to the 
$K$-th derivative of the $N$ generalized coordinates $q_i$, i.e. 
\[
L\left( q_i,\stackrel{.}{q}_i,...,\stackrel{\left( K\right) }{q}_i\right) ;\ 
\stackrel{\left( s\right) }{q}_i\equiv \frac{d^sq_i}{dt^s}, 
\]
were $s=0,1,...,K$ and $i=1,...N$. For such systems the Euler-Lagrange
equations of motion, obtained through Hamilton's principle of stationary
action, will be 
\[
\stackrel{K}{\stackunder{s=0}{\sum }}\left( -1\right) ^s\frac{d^s}{dt^s}%
\left( \frac{\partial L}{\partial \stackrel{\left( s\right) }{q}_i}\right)
=0. 
\]

This is a system of 
$N$ differential equations of $2K$-th order so we need $2KN$ initial
conditions to solve it. These conditions are the initial values of $q_i,%
\stackrel{.}{q}_i,...,\stackrel{\left( 2K-1\right) }{q_i}$ that describe the 
{\it velocity} phase space ({\it VPS}).

The Hamiltonian formalism for theories with higher order derivatives, that
has been first developed by Ostrogradski \cite{Ostrogradski}, treats the
derivatives $\stackrel{\left( s\right) }{q}$ ($s=0,...,K-1$) as coordinates.
So we will indicate this writing them as $\stackrel{\left( s\right) }{q}%
_i\equiv q_{\left( s\right) i}$. In Ostrogradski's formalism the momenta
conjugated respectively to $q_{\left( K-1\right) i}$ and $q_{\left(
s-1\right) i}$ ($s=1,...,K-1$) are introduced as 
\begin{equation}
p_{\left( K-1\right) i}\equiv \frac{\partial L}{\partial \stackrel{\left(
K\right) }{q}_i},  \label{Eq3}
\end{equation}
\begin{equation}
p_{\left( s-1\right) i}\equiv \frac{\partial L}{\partial \stackrel{\left(
s\right) }{q}_i}-\stackrel{.}{p}_{\left( s\right) i};\ s=1,...,K-1.
\label{Eq4}
\end{equation}
Notice that the momenta $p_{\left( s\right) i}$ ($s\geq 0$) will only be
dependent on the derivatives up to $\stackrel{\left( 2K-1-s\right) }{q_i}$.

The Hamiltonian is defined as 
\begin{equation}
H=\stackrel{K-1}{\stackunder{s=0}{\sum }}p_{\left( s\right) i}\stackrel{%
\left( s+1\right) }{q_i}-L\left( q_i,...,\stackrel{\left( K\right) }{q}%
_i\right) ,  \label{Eq5}
\end{equation}
where we use Einstein's summation rule for repeated indexes as will be done
throughout this paper. Anyway, we will write explicitly the summation over
the index $\left( s\right) $ inside the parenthesis for a question of
intelligibility.

The Hamilton's equations of motion will be written as 
\begin{equation}
\stackrel{.}{q}_{\left( s\right) i}=\frac{\partial H}{\partial p_{\left(
s\right) i}}=\left\{ q_{\left( s\right) i},H\right\} ,  \label{Eq6}
\end{equation}
\begin{equation}
\stackrel{.}{p}_{\left( s\right) i}=-\frac{\partial H}{\partial q_{\left(
s\right) i}}=\left\{ p_{\left( s\right) i},H\right\} ,  \label{Eq7}
\end{equation}
were $\left\{ \ ,\ \right\} $ is the Poisson bracket defined as 
\[
\left\{ A,B\right\} \equiv \stackrel{K-1}{\stackunder{s=0}{\sum }}\frac{%
\partial A}{\partial q_{\left( s\right) i}}\frac{\partial B}{\partial
p_{\left( s\right) i}}-\frac{\partial B}{\partial q_{\left( s\right) i}}%
\frac{\partial A}{\partial p_{\left( s\right) i}}. 
\]

The fundamental Poisson brackets are 
\[
\left\{ q_{\left( s\right) i},p_{\left( s^{\prime }\right) j}\right\}
=\delta _{ss^{\prime }}\delta _{ij};\ \left\{ q_{\left( s\right)
i},q_{\left( s^{\prime }\right) j}\right\} =\left\{ p_{\left( s\right)
i},p_{\left( s^{\prime }\right) j}\right\} =0, 
\]
were $i,j=1,...,N$ and $s,s^{\prime }=0,...,K-1$. With this procedure the
phase space ({\it PS}) is described in terms of the canonical variables $%
q_{\left( s\right) i}$ and $p_{\left( s\right) i}$ (where $i=1,...,N$ and $%
s=0,...,K-1$) obeying $2NK$ equations of motion (given by equations (\ref
{Eq6}) and (\ref{Eq7})) which are first order differential equations. So we
have to fix $2KN$ initial conditions to solve these equations. These initial
conditions are analogue to those needed in Euler-Lagrange equations, but now
they are the initial values of the canonical variables.

However, this passage from {\it VPS} to {\it PS} is only possible if we can
solve the momenta expressions (\ref{Eq3}) and (\ref{Eq4}) with respect to
the derivatives $\stackrel{\left( K\right) }{q}_i,...,\stackrel{\left(
2K-1\right) }{q_i}$ so that these can be expressed as functions of the
canonical variables and eliminated from the theory. The necessity of
expressing the derivatives $\stackrel{\left( K\right) }{q}_i$ as functions
of the canonical variables is clear since these derivatives are present in
the Hamiltonian definition (\ref{Eq5}) and must be eliminated from all
equations in the Hamiltonian formulation. This procedure is completely
analogous to the elimination of the velocities $\stackrel{.}{q}_i$ in the
Hamiltonian formalism of a first order system \cite{Sundermeyer, Mukunda}.

The same argument can not be applied to the derivatives 
$\stackrel{\left( K+1\right) }{q_i},...,\stackrel{\left( 2K-1\right) }{q_i}$
since derivatives higher than $\stackrel{\left( K\right) }{q}_i$ are not
present in the Hamiltonian. But now, the necessity of expressing these
derivatives as functions of canonical variables comes from the fact that
they are present in the momenta expressions (\ref{Eq4}). So, fixing the
initial conditions of these momenta in the Hamiltonian formulation is
equivalent to fixing the initial conditions to the derivatives $\stackrel{%
\left( K+1\right) }{q_i},...,\stackrel{\left( 2K-1\right) }{q_i}$ in the
Lagrangian formulation. The same relation connects the momenta $p_{\left(
K-1\right) i}$ and the derivative $\stackrel{\left( K\right) }{q}_i$: fixing
the initial conditions for the momenta $p_{\left( K-1\right) i}$ in the
Hamiltonian formulation is equivalent to fixing the initial condition for
the derivatives $\stackrel{\left( K\right) }{q}_i$. Then, it is necessary
that all the momenta (\ref{Eq3}) and (\ref{Eq4}) be linearly independent
functions of the derivatives $\stackrel{\left( K\right) }{q}_i,...,\stackrel{%
\left( 2K-1\right) }{q_i}$ so that the latter ones can be solved uniquely
with respect to the former.

The expression (\ref{Eq3}) for the momenta $p_{\left( K-1\right) i}$ shows
that they are dependent only on derivatives up to $\stackrel{\left( K\right) 
}{q}_i$, so these derivatives can be solved as functions 
\begin{equation}
\stackrel{\left( K\right) }{q}_i=f_{\left( K\right) i}\left( q_{\left(
s\right) j};\ p_{\left( K-1\right) j}\right) ;\ s=0,...,K-1,  \label{Eq10}
\end{equation}
if, and only if, the momenta $p_{\left( K-1\right) i}$ are linearly
independent functions of the derivatives $\stackrel{\left( K\right) }{q}_i$.
For this, it is necessary that the matrix 
\begin{equation}
H_{ij}\equiv \frac{\partial p_{\left( K-1\right) i}}{\partial \stackrel{%
\left( K\right) }{q}_j}=\frac{\partial ^2L}{\partial \stackrel{\left(
K\right) }{q}_i\partial \stackrel{\left( K\right) }{q}_j}  \label{Eq11}
\end{equation}
be non singular. This matrix $H_{ij}$ is called the {\it Hessian} matrix of
the system and is simply the Jacobian matrix of the change of variables $%
\stackrel{\left( K\right) }{q}_j\rightarrow p_{\left( K-1\right) i}$.

Now, from definitions (\ref{Eq3}) and (\ref{Eq4}), we can see that the
momenta $p_{\left( K-2\right) i}$ are dependent on derivatives up to $%
\stackrel{\left( K+1\right) }{q_i}$. In addition, the dependence on $%
\stackrel{\left( K+1\right) }{q_i}$ is linear and the coefficients are the
elements $H_{ij}$ of the Hessian matrix. Considering that the derivatives $%
\stackrel{\left( K\right) }{q}_i$ have already been eliminated from
equations using the expression (\ref{Eq10}) the derivatives $\stackrel{%
\left( K+1\right) }{q_i}$ can be solved as functions 
\[
\stackrel{\left( K+1\right) }{q_i}=f_{\left( K+1\right) i}\left( q_{\left(
s\right) j};\ p_{\left( K-1\right) j},\ p_{\left( K-2\right) j}\right) ;\
s=0,...,K-1, 
\]
if, and only if, the momenta $p_{\left( K-2\right) i}$ are linearly
independent functions of the derivatives $\stackrel{\left( K+1\right) }{q_i}$%
. For that it will be necessary that the Jacobian matrix of the change of
variables $\stackrel{\left( K+1\right) }{q_j}\rightarrow p_{\left(
K-2\right) i}$, with elements $J_{ij}$ given by 
\[
J_{ij}=\left. \frac{\partial p_{\left( K-2\right) i}}{\partial \stackrel{%
\left( K+1\right) }{q_j}}\right| _{\stackrel{\left( K\right) }{q}%
_u=f_{\left( K\right) u}}=-\left. \frac{\partial ^2L}{\partial \stackrel{%
\left( K\right) }{q}_i\partial \stackrel{\left( K\right) }{q}_j}\right| _{%
\stackrel{\left( K\right) }{q}_u=f_{\left( K\right) u}}=-\left.
H_{ij}\right| _{\stackrel{\left( K\right) }{q}_u=f_{\left( K\right) u}}, 
\]
be non singular.

Continuing this process, we can use the fact that the momenta 
$p_{\left( s\right) i}$ are dependent on derivatives up to $\stackrel{\left(
2K-1-s\right) }{q_i}$ and, noticing that the highest derivatives appear
linearly with coefficients that are the elements of the Hessian matrix, show
that the derivatives $\stackrel{\left( K+p\right) }{q_i}$ can be solved as
functions 
\begin{equation}
\stackrel{\left( K+p\right) }{q_i}=f_{\left( K+p\right) i}\left( q_{\left(
s\right) j};\ p_{\left( K-1\right) j},...,\ p_{\left( K-1-p\right) j}\right)
;\ s,p=0,...,K-1  \label{Eq14}
\end{equation}
if, and only if, the Jacobian matrix of the change of variables $\stackrel{%
\left( K+p\right) }{q_j}\rightarrow p_{\left( K-1-p\right) i}$, with
elements $J_{ij}$ given by 
\begin{eqnarray*}
J_{ij} &=&\left. \frac{\partial p_{\left( K-1-p\right) i}}{\partial 
\stackrel{\left( K+p\right) }{q_j}}\right| _{\stackrel{\left( K+d\right) }{q}%
_u=f_{\left( K+d\right) u}}=\left( -1\right) ^p\left. \frac{\partial ^2L}{%
\partial \stackrel{\left( K\right) }{q}_i\partial \stackrel{\left( K\right) 
}{q}_j}\right| _{\stackrel{\left( K+d\right) }{q}_u=f_{\left( K+d\right) u}}
\\
&& \\
J_{ij} &=&\left( -1\right) ^p\left. H_{ij}\right| _{\stackrel{\left(
K+d\right) }{q}_u=f_{\left( K+d\right) u}}
\end{eqnarray*}
($d=0,...,p-1$), is non singular. Consequently, it will be the non
singularity of the Hessian matrix (\ref{Eq11}) that will determine if the
passage from the {\it VPS} to {\it PS} is possible or not.

Let's suppose now that the Hessian matrix has rank 
$P=N-R$. In this case it will not be possible to express all derivatives $%
\stackrel{\left( K\right) }{q}_i,...,\stackrel{\left( 2K-1\right) }{q_i}$ in
the form of equation (\ref{Eq14}). Without loss of generality, we can choose
the order of coordinates in such a way that the $P\times P$ sub-matrix in
the bottom right corner of the Hessian matrix has nonvanishing determinant 
\[
\det \left\| H_{ab}\right\| =\det \left\| \frac{\partial ^2L}{\partial 
\stackrel{\left( K\right) }{q}_a\partial \stackrel{\left( K\right) }{q}_b}%
\right\| \neq 0;\ a,b=R+1,...,N. 
\]

With this condition, we can only solve the 
$P=N-R$ derivatives $\stackrel{\left( K\right) }{q}_a$ as functions of the
coordinates $q_{\left( s\right) i}$, the momenta $p_{\left( K-1\right) b}$
and the unsolved derivatives $\stackrel{\left( K\right) }{q}_\alpha $ ($%
\alpha =1,...,R$) as follows 
\[
\stackrel{\left( K\right) }{q}_a=f_{\left( K\right) a}\left( q_{\left(
s\right) i};\ p_{\left( K-1\right) b};\stackrel{\left( K\right) }{q}_\alpha
\right) . 
\]

If we substitute this expression in the momenta definition (\ref{Eq3}) for $%
p_{\left( K-1\right) i}$ we obtain 
\[
p_{\left( K-1\right) i}=\left. \frac{\partial L}{\partial \stackrel{\left(
K\right) }{q}_i}\right| _{\stackrel{\left( K\right) }{q}_a=f_{\left(
K\right) a}}=g_{\left( K-1\right) i}\left( q_{\left( s\right) j};\ p_{\left(
K-1\right) a};\stackrel{\left( K\right) }{q}_\alpha \right) . 
\]

But, since we have 
$p_{\left( K-1\right) a}\equiv g_{\left( K-1\right) a}$, the other $R$
functions $g_{\left( K-1\right) \alpha }$ can not contain the unsolved
derivatives $\stackrel{\left( K\right) }{q}_\alpha $ or we would be able to
solve more of these derivatives as functions of the canonical variables,
what contradicts the fact that the rank of the Hessian matrix is $P$. So we
have the expressions 
\begin{equation}
p_{\left( K-1\right) \alpha }=g_{\left( K-1\right) \alpha }\left( q_{\left(
s\right) j};\ p_{\left( K-1\right) a}\right)  \label{Eq19}
\end{equation}
which correspond to primary constraints 
\begin{equation}
\Phi _{\left( K-1\right) \alpha }=p_{\left( K-1\right) \alpha }-g_{\left(
K-1\right) \alpha }\left( q_{\left( s\right) j};\ p_{\left( K-1\right)
a}\right) \approx 0  \label{Eq20}
\end{equation}
in Dirac's Hamiltonian formalism for singular systems.

Analogously, we can only solve the 
$P$ derivatives $\stackrel{\left( K+1\right) }{q}_a$ as functions of the
coordinates $q_{\left( s\right) i}$, the momenta $p_{\left( K-1\right) b}$
and $\ p_{\left( K-2\right) b}$ and the unsolved derivatives $\stackrel{%
\left( K\right) }{q}_\alpha $ and $\stackrel{\left( K+1\right) }{q_\alpha }$
($\alpha =1,...,R$) as follows 
\[
\stackrel{\left( K+1\right) }{q_a}=f_{\left( K+1\right) a}\left( q_{\left(
s\right) i};\ p_{\left( K-1\right) b},\ \ p_{\left( K-2\right) b};\stackrel{%
\left( K\right) }{q}_\alpha ,\ \stackrel{\left( K+1\right) }{q_\alpha }%
\right) . 
\]

Substituting this expression in momenta definitions (\ref{Eq4}) and using
the above argument relative to the rank of the Hessian matrix, we obtain for
the momenta $p_{\left( K-2\right) \alpha }$ the following expression 
\[
p_{\left( K-2\right) \alpha }=g_{\left( K-2\right) \alpha }\left( q_{\left(
s\right) j};\ p_{\left( K-1\right) a},\ p_{\left( K-2\right) a}\right) 
\]
that corresponds to new primary constraints 
\[
\Phi _{\left( K-2\right) \alpha }=p_{\left( K-2\right) \alpha }-g_{\left(
K-2\right) \alpha }\left( q_{\left( s\right) j};\ p_{\left( K-1\right) a},\
p_{\left( K-2\right) a}\right) \approx 0. 
\]

Continuing this process we find that there will be expressions 
\begin{equation}
p_{\left( K-1-p\right) \alpha }=g_{\left( K-1-p\right) \alpha }\left(
q_{\left( s\right) j};\ p_{\left( K-1\right) a},...,\ p_{\left( K-1-p\right)
a}\right)  \label{Eq24}
\end{equation}
($p=0,...,K-1$) that correspond to primary constraints 
\begin{equation}
\Phi _{\left( K-1-p\right) \alpha }=p_{\left( K-1-p\right) \alpha
}-g_{\left( K-1-p\right) \alpha }\left( q_{\left( s\right) j};\ p_{\left(
K-1\right) a},...,\ p_{\left( K-1-p\right) a}\right) \approx 0.  \label{Eq25}
\end{equation}

As a result, in a higher order system, the existence of constraints
involving a given momentum 
$p_{\left( K-1\right) \alpha }$ will imply the existence of constraints
involving all $p_{\left( s\right) \alpha }$ momenta conjugated to the
derivatives $q_{\left( s\right) \alpha }=\stackrel{\left( s\right) }{q}%
_\alpha $ ($s=0,...,K-1$) due to the fact that the derivatives $\stackrel{%
\left( K\right) }{q}_\alpha ,...,\stackrel{\left( 2K-1\right) }{q_\alpha }$
can't be expressed as functions of the canonical variables. Consequently, if
the Hessian matrix has rank $P=N-R$ there are $KR$ expressions of the form (%
\ref{Eq24}) that correspond to $KR$ primary constraints in Dirac's formalism
as given by (\ref{Eq25}).

The existence of such {\it constraints' structure} in higher order systems
has already been noticed by other authors. Nesterenko \cite{Nesterenko} and
Batlle {\it et al.} \cite{Batlle} discerned the existence of such
constraints structure in second order systems, while Saito {\it et al.} \cite
{Saito} showed, by different arguments, that the constraints structure
exhibited above exists for arbitrarily higher order systems. Furthermore, it
is important to observe that the constraint structure showed above is
different for a higher order Lagrangian obtained from a lower order one by
adding a total time derivative. We will not discuss this case here but the
reader can find a detailed analysis of the constraint structure for such
Lagrangians in reference \cite{Saito}.

\section{Hamilton-Jacobi formalism}

Now we will use Carath\'{e}odory's equivalent Lagrangians method to extend
the Hamilton-Jacobi for\-ma\-lism to a general higher order Lagrangian. The
procedure described in the sequence can be applied to any higher order
Lagrangian and is not restricted to a singular one.

Carath\'{e}odory's equivalent Lagrangians method \cite{Caratheodory} can be
easily applied to higher order Lagrangians. Given a Lagrangian $L\left( q_i,%
\stackrel{.}{q}_i,...,\stackrel{\left( K\right) }{q}_i\right) $, we can
obtain a completely equivalent one given by 
\[
L^{\prime }=L\left( q_i,...,\stackrel{\left( K\right) }{q}_i\right) -\frac{%
dS\left( q_i,...,q_{\left( K-1\right) i},t\right) }{dt}. 
\]

These Lagrangians are equivalent because the action integral given by them
have simultaneous extremum. So we can choose the function 
$S\left( q_i,...,q_{\left( K-1\right) i},t\right) $ in such a way that we
get an extremum of $L^{\prime }$ and consequently we will get an extremum of
the Lagrangian $L$.

To do this, it is enough to find a set of functions 
$\beta _{\left( s\right) i}(q_j,q_{\left( 1\right) j},...,q_{\left(
s-1\right) j},t)$, $s=1,...,K$, and $S\left( q_i,...,q_{\left( K-1\right)
i},t\right) $ such that 
\begin{equation}
L^{\prime }\left( q_i,q_{\left( 1\right) i}=\beta _{\left( 1\right)
i},...,q_{\left( K\right) i}=\beta _{\left( K\right) i},t\right) =0
\label{Eq.15}
\end{equation}
and for all neighborhood of $q_{\left( s\right) i}=\beta _{\left( s\right)
i}(q_j,q_{\left( 1\right) j},...,q_{\left( s-1\right) j},t)$ 
\begin{equation}
L^{\prime }\left( q_i,...,\stackrel{\left( K\right) }{q}_i\right) >0.
\label{Eq.16}
\end{equation}

With these conditions satisfied, the Lagrangian 
$L^{\prime }$ will have a minimum in $q_{\left( s\right) i}=\beta _{\left(
s\right) i}$ so that the action integral will also have a minimum and the
solutions of the differential equations given by 
\[
q_{\left( s\right) i}=\stackrel{\left( s\right) }{q}_i=\frac{d^sq_i}{dt^s}%
=\beta _{\left( s\right) i}, 
\]
$s=1,...,K$, will correspond to an extremum of the action integral.

From the definition of 
$L^{\prime }$ we have 
\[
L^{\prime }=L\left( q_j,...,\stackrel{\left( K\right) }{q}_j\right) -\frac{%
\partial S\left( q_j,...,q_{\left( K-1\right) j},t\right) }{\partial t}-%
\stackrel{K-1}{\stackunder{u=0}{\sum }}\frac{\partial S\left(
q_j,...,q_{\left( K-1\right) j},t\right) }{\partial q_{\left( u\right) i}}%
\frac{dq_{\left( u\right) i}}{dt}. 
\]

Using condition (\ref{Eq.15}) we obtain 
\begin{eqnarray*}
&&\left[ L\left( q_j,...,\stackrel{\left( K\right) }{q}_j\right) -\frac{%
\partial S\left( q_j,...,q_{\left( K-1\right) j},t\right) }{\partial t}%
\right. \\
&&_{}{}\hskip 2.8cm\left. \left. -\stackrel{K-1}{\stackunder{u=0}{\sum }}%
\frac{\partial S\left( q_j,...,q_{\left( K-1\right) j},t\right) }{\partial
q_{\left( u\right) i}}q_{\left( u+1\right) i}\right] \right| _{q_{\left(
s\right) i}=\beta _{\left( s\right) i}}=0, \\
&&
\end{eqnarray*}
\begin{equation}
\left. \frac{\partial S}{\partial t}\right| _{q_{\left( s\right) i}=\beta
_{\left( s\right) i}}=\left. \left[ L\left( q_j,...,\stackrel{\left(
K\right) }{q}_j\right) -\stackrel{K-1}{\stackunder{u=0}{\sum }}\frac{%
\partial S\left( q_j,...,q_{\left( K-1\right) j},t\right) }{\partial
q_{\left( u\right) i}}q_{\left( u+1\right) i}\right] \right| _{q_{\left(
s\right) i}=\beta _{\left( s\right) i}}.  \label{Eq.19}
\end{equation}

Since 
$q_{\left( s\right) i}=\beta _{\left( s\right) i}$ is a minimum point of $%
L^{\prime }$ we must have 
\[
\left. \frac{\partial L^{\prime }}{\partial \stackrel{\left( K\right) }{q}_i}%
\right| _{q_{\left( s\right) i}=\beta _{\left( s\right) i}}=0\Rightarrow
\left. \left[ \frac{\partial L}{\partial \stackrel{\left( K\right) }{q}_i}%
-\frac \partial {\partial \stackrel{\left( K\right) }{q}_i}\left( \frac{dS}{%
dt}\right) \right] \right| _{q_{\left( s\right) i}=\beta _{\left( s\right)
i}}=0, 
\]
\[
\left. \left[ \frac{\partial L}{\partial \stackrel{\left( K\right) }{q}_i}-%
\frac{\partial S}{\partial q_{\left( K-1\right) i}}\right] \right|
_{q_{\left( s\right) i}=\beta _{\left( s\right) i}}=0, 
\]
or 
\begin{equation}
\left. \frac{\partial S}{\partial q_{\left( K-1\right) i}}\right|
_{q_{\left( s\right) i}=\beta _{\left( s\right) i}}=\left. \frac{\partial L}{%
\partial \stackrel{\left( K\right) }{q}_i}\right| _{q_{\left( s\right)
i}=\beta _{\left( s\right) i}}.  \label{Eq.22}
\end{equation}

For the same reason we must have 
\[
\left. \frac{\partial L^{\prime }}{\partial \stackrel{\left( K-1\right) }{q_i%
}}\right| _{q_{\left( s\right) i}=\beta _{\left( s\right) i}}=0\Rightarrow
\left. \left[ \frac{\partial L}{\partial q_{\left( K-1\right) i}}-\frac
\partial {\partial q_{\left( K-1\right) i}}\left( \frac{dS}{dt}\right)
\right] \right| _{q_{\left( s\right) i}=\beta _{\left( s\right) i}}=0, 
\]
\[
\left. \left[ \frac{\partial L}{\partial q_{\left( K-1\right) i}}-\frac
\partial {\partial t}\frac{\partial S}{\partial q_{\left( K-1\right) i}}-%
\frac{\partial S}{\partial q_{\left( K-2\right) i}}-\stackrel{K-1}{%
\stackunder{u=0}{\sum }}\frac{\partial ^2S}{\partial q_{\left( K-1\right)
i}\partial q_{\left( u\right) j}}\frac{dq_{\left( u\right) j}}{dt}\right]
\right| _{q_{\left( s\right) i}=\beta _{\left( s\right) i}}=0, 
\]
\[
\left. \left[ \frac{\partial L}{\partial q_{\left( K-1\right) i}}-\frac{%
\partial S}{\partial q_{\left( K-2\right) i}}-\frac d{dt}\frac{\partial S}{%
\partial q_{\left( K-1\right) i}}\right] \right| _{q_{\left( s\right)
i}=\beta _{\left( s\right) i}}=0, 
\]
or 
\begin{equation}
\left. \frac{\partial S}{\partial q_{\left( K-2\right) i}}\right|
_{q_{\left( s\right) i}=\beta _{\left( s\right) i}}=\left. \left[ \frac{%
\partial L}{\partial q_{\left( K-1\right) i}}-\frac d{dt}\frac{\partial S}{%
\partial q_{\left( K-1\right) i}}\right] \right| _{q_{\left( s\right)
i}=\beta _{\left( s\right) i}}.  \label{Eq.26}
\end{equation}

Following this procedure we have the general expression 
\begin{equation}
\left. \frac{\partial S}{\partial q_{\left( u-1\right) i}}\right|
_{q_{\left( s\right) i}=\beta _{\left( s\right) i}}=\left. \left[ \frac{%
\partial L}{\partial q_{\left( u\right) i}}-\frac d{dt}\frac{\partial S}{%
\partial q_{\left( u\right) i}}\right] \right| _{q_{\left( s\right) i}=\beta
_{\left( s\right) i}}  \label{Eq.26b}
\end{equation}
were $u=1,...,K-1$.

Now, using the definitions for the conjugated momenta given by equations (%
\ref{Eq3}) and (\ref{Eq4}) in the expressions (\ref{Eq.22}) and (\ref{Eq.26b}%
) we obtain 
\begin{equation}
p_{\left( u\right) i}=\frac{\partial S}{\partial q_{\left( u\right) i}},\
u=0,...,K-1.  \label{Eq.29}
\end{equation}

So, we can see from equation (\ref{Eq.19}) that, to obtain an extremum of
the action, we must get a function $S\left( q_i,...,q_{\left( K-1\right)
i},t\right) $ such that 
\begin{equation}
\frac{\partial S}{\partial t}=-H_0  \label{Eq.27}
\end{equation}
where $H_0$ is 
\begin{equation}
H_0=\stackrel{K-1}{\stackunder{u=0}{\sum }}p_{\left( u\right) i}\stackrel{%
\left( u+1\right) }{q_i}-L\left( q_i,...,\stackrel{\left( K\right) }{q}%
_i\right)  \label{Eq.28}
\end{equation}
and the momenta $p_{\left( u\right) i}$ are given by equation (\ref{Eq.29}).

These are the fundamental equations of the equivalent Lagrangian method, and
equation (\ref{Eq.27}) is the Hamilton-Jacobi partial differential equation
(HJPDE).

\section{The singular case}

We consider now the application of the formalism developed in the previous
section to a system with a singular higher order Lagrangian. As we showed in
Sect. 2 , when the Hessian matrix has a rank 
$P=N-R$ the momenta variables will not be independent among themselves and
we will obtain expressions like equation (\ref{Eq24}). We will rewrite these
expressions as 
\begin{equation}
p_{\left( u\right) \alpha }=-H_{\left( u\right) \alpha }\left( q_{\left(
s\right) j};\ p_{\left( s\right) a}\right) ;\ u,\ s=0,...,K-1  \label{Eq.29b}
\end{equation}
where we are supposing that the expression for the momentum $p_{\left(
u\right) \alpha }$ depends on all momenta $p_{\left( s\right) a}$, although
we have showed that the expression for the momentum $p_{\left( u\right)
\alpha }$ is not dependent on any momenta $p_{\left( s\right) \alpha }$ with %
$s<u$. We do this for simplicity.

The Hamiltonian 
$H_0$, given by equation (\ref{Eq.28}), becomes 
\begin{eqnarray}
H_0 &=&\stackrel{K-2}{\stackunder{u=0}{\sum }}p_{\left( u\right) a}\stackrel{%
\left( u+1\right) }{q_a}+p_{\left( K-1\right) a}f_{\left( K\right) a}+%
\stackrel{K-1}{\stackunder{u=0}{\sum }}\stackrel{\left( u+1\right) }{%
q_\alpha }\left. p_{\left( u\right) \alpha }\right| _{p_{\left( s\right)
\beta }=-H_{\left( s\right) \beta }}{}  \nonumber \\
&&-L\left( q_{\left( s\right) i},\stackrel{\left( K\right) }{q}_\alpha ,%
\stackrel{\left( K\right) }{q}_a=f_{\left( K\right) a}\right) ,
\label{Eq.36}
\end{eqnarray}
where $\alpha ,\beta =1,...,R$; $a=R+1,...,N$. On the other hand we have 
\[
\frac{\partial H_0}{\partial \stackrel{\left( K\right) }{q}_\alpha }%
=p_{\left( K-1\right) a}\frac{\partial f_{\left( K\right) a}}{\partial 
\stackrel{\left( K\right) }{q}_\alpha }+p_{\left( K-1\right) \alpha }-\frac{%
\partial L}{\partial \stackrel{\left( K\right) }{q}_\alpha }-\frac{\partial L%
}{\partial \stackrel{\left( K\right) }{q}_\alpha }\frac{\partial f_{\left(
K\right) a}}{\partial \stackrel{\left( K\right) }{q}_\alpha }=0, 
\]
so the Hamiltonian $H_0$ does not depend explicitly upon the derivatives $%
\stackrel{\left( K\right) }{q}_\alpha $.

Now we will adopt the following notation: the time parameter 
$t$ will be called $t_{\left( s\right) 0}\equiv q_{\left( s\right) 0}$ (for
any value of $s$); the coordinates $q_{\left( s\right) \alpha }$ will be
called $t_{\left( s\right) \alpha }$; the momenta $p_{\left( s\right) \alpha
}$ will be called $P_{\left( s\right) \alpha }$ and the {\it momentum} $%
p_{\left( s\right) 0}\equiv P_{\left( s\right) 0}$ will be defined as 
\begin{equation}
P_{\left( s\right) 0}\equiv \frac{\partial S}{\partial t},  \label{Eq.38}
\end{equation}
while $H_{\left( s\right) 0}\equiv H_0$ for any value of $s$.

Then, to obtain an extremum of the action integral, we must find a function 
$S\left( t_{\left( c\right) \alpha };q_{\left( c\right) a},t\right) $ $%
\left( c=0,...,K-1\right) $ that satisfies the following set of HJPDE 
\begin{equation}
H_0^{\prime }\equiv H_{\left( s\right) 0}^{\prime }\equiv P_{\left( s\right)
0}+H_{\left( s\right) 0}\left( t,t_{\left( u\right) \alpha };q_{\left(
u\right) a};p_{\left( u\right) a}=\frac{\partial S}{\partial q_{\left(
u\right) a}}\right) =0,  \label{Eq.39}
\end{equation}
\begin{equation}
H_{\left( s\right) \alpha }^{\prime }\equiv P_{\left( s\right) \alpha
}+H_{\left( s\right) \alpha }\left( t_{\left( u\right) \alpha };q_{\left(
u\right) a};p_{\left( u\right) a}=\frac{\partial S}{\partial q_{\left(
u\right) a}}\right) =0.  \label{Eq.40}
\end{equation}
where $s,u=0,...,K-1$ and $\alpha =1,...,R$. If we let the index $\alpha $
run from $0$ to $R$ we can write both equations as 
\begin{equation}
H_{\left( s\right) \alpha }^{\prime }\equiv P_{\left( s\right) \alpha
}+H_{\left( s\right) \alpha }\left( t_{\left( u\right) \alpha };q_{\left(
u\right) a};p_{\left( u\right) a}=\frac{\partial S}{\partial q_{\left(
u\right) a}}\right) =0.  \label{Eq.41}
\end{equation}

From the above definition above and equation (\ref{Eq.36}) we have 
\[
\frac{\partial H_{\left( s\right) 0}^{\prime }}{\partial p_{\left( u\right)
b}}=-\frac{\partial L}{\partial \stackrel{\left( K\right) }{q}_a}\frac{%
\partial f_{\left( K\right) a}}{\partial p_{\left( u\right) b}}-\stackrel{K-1%
}{\stackunder{s=0}{\sum }}\stackrel{\left( s+1\right) }{q}_\alpha \frac{%
\partial H_{\left( s\right) \alpha }}{\partial p_{\left( u\right) b}}%
+p_{\left( K-1\right) a}\frac{\partial f_{\left( K\right) a}}{\partial
p_{\left( u\right) b}}+q_{\left( u+1\right) b}, 
\]
\[
\frac{\partial H_{\left( s\right) 0}^{\prime }}{\partial p_{\left( u\right)
b}}=\stackrel{.}{q}_{\left( u\right) b}-\stackrel{K-1}{\stackunder{s=0}{\sum 
}}\stackrel{.}{q}_{\left( s\right) \alpha }\frac{\partial H_{\left( s\right)
\alpha }}{\partial p_{\left( u\right) b}}, 
\]
where $u=0,...,K-1$, $\alpha =1,...,R$ and we used the fact that 
\[
\stackrel{\left( c+1\right) }{q_i}=q_{\left( c+1\right) i}=\frac{dq_{\left(
c\right) i}}{dt};\ c=0,...,K-1. 
\]

Multiplying this equation by 
$dt=dt_{\left( s\right) 0}$ we have 
\[
dq_{\left( u\right) b}=\frac{\partial H_{\left( s\right) 0}^{\prime }}{%
\partial p_{\left( u\right) b}}dt_{\left( s\right) 0}+\stackrel{K-1}{%
\stackunder{s=0}{\sum }}\frac{\partial H_{\left( s\right) \alpha }^{\prime }%
}{\partial p_{\left( u\right) b}}dq_{\left( s\right) \alpha }. 
\]

Using 
$t_{\left( s\right) \alpha }=q_{\left( s\right) \alpha }$ and making the
index $\alpha $ run from $0$ to $R$, we have 
\begin{equation}
dq_{\left( u\right) b}=\stackrel{K-1}{\stackunder{s=0}{\sum }}\frac{\partial
H_{\left( s\right) \alpha }^{\prime }}{\partial p_{\left( u\right) b}}%
dt_{\left( s\right) \alpha }.  \label{Eq.46}
\end{equation}

We must call attention to the fact that in the above expression, for 
$\alpha =0$, we have the term 
\[
\stackrel{K-1}{\stackunder{s=0}{\sum }}\frac{\partial H_{\left( s\right)
0}^{\prime }}{\partial p_{\left( u\right) b}}dt_{\left( s\right) 0}\equiv 
\frac{\partial H_0^{\prime }}{\partial p_{\left( u\right) b}}dt 
\]
that should not be interpreted as 
\[
\stackrel{K-1}{\stackunder{s=0}{\sum }}\frac{\partial H_{\left( s\right)
0}^{\prime }}{\partial p_{\left( u\right) b}}dt_{\left( s\right) 0}=K\cdot 
\frac{\partial H_0^{\prime }}{\partial p_{\left( u\right) b}}dt. 
\]
This somewhat unusual choice of notation allows us to express the results in
a compact way.

Noticing that we have the expressions 
\[
dq_{\left( u\right) \beta }=\stackrel{K-1}{\stackunder{s=0}{\sum }}\frac{%
\partial H_{\left( s\right) \alpha }^{\prime }}{\partial p_{\left( u\right)
\beta }}dt_{\left( s\right) \alpha }=\delta _{su}\delta _{\alpha \beta
}dt_{\left( s\right) \alpha }\equiv dt_{\left( s\right) \beta } 
\]
identically satisfied for $\alpha ,\ \beta =0,1,...,R$, we can write the
expression (\ref{Eq.46}) as 
\begin{equation}
dq_{\left( u\right) i}=\stackrel{K-1}{\stackunder{s=0}{\sum }}\frac{\partial
H_{\left( s\right) \alpha }^{\prime }}{\partial p_{\left( u\right) i}}%
dt_{\left( s\right) \alpha };\ i=1,...,N.  \label{Eq.50}
\end{equation}

If we consider that we have a solution 
$S\left( q_i,...,q_{\left( K-1\right) i},t\right) $ of the set of HJPDE
given by equation (\ref{Eq.41}) then, differentiating that equation with
respect to $q_{\left( u\right) c}$, we obtain 
\begin{equation}
\frac{\partial H_{\left( s\right) \alpha }^{\prime }}{\partial q_{\left(
u\right) c}}+\stackrel{K-1}{\stackunder{d=0}{\sum }}\frac{\partial H_{\left(
s\right) \alpha }^{\prime }}{\partial P_{\left( d\right) \beta }}\frac{%
\partial ^2S}{\partial t_{\left( d\right) \beta }\partial q_{\left( u\right)
c}}+\stackrel{K-1}{\stackunder{d=0}{\sum }}\frac{\partial H_{\left( s\right)
\alpha }^{\prime }}{\partial p_{\left( d\right) a}}\frac{\partial ^2S}{%
\partial q_{\left( d\right) a}\partial q_{\left( u\right) c}}=0
\label{Eq.52}
\end{equation}
for $\alpha ,\ \beta =0,1,...,R$; $s,\ u,\ d=0,1,...,K-1$ and $c=0,1,...,N$.

From the momenta definitions we can obtain 
\begin{equation}
dp_{\left( u\right) c}=\stackrel{K-1}{\stackunder{d=0}{\sum }}\frac{\partial
^2S}{\partial q_{\left( u\right) c}\partial t_{\left( d\right) \beta }}%
dt_{\left( d\right) \beta }+\stackrel{K-1}{\stackunder{d=0}{\sum }}\frac{%
\partial ^2S}{\partial q_{\left( u\right) c}\partial q_{\left( d\right) a}}%
dq_{\left( d\right) a}.  \label{Eq.65}
\end{equation}

Now, contracting equation (\ref{Eq.52}) with $dt_{\left( s\right) \alpha }$
and adding the result to equation (\ref{Eq.65}) we get 
\begin{eqnarray*}
dp_{\left( u\right) c}+\stackrel{K-1}{\stackunder{s=0}{\sum }}\frac{\partial
H_{\left( s\right) \alpha }^{\prime }}{\partial q_{\left( u\right) c}}%
dt_{\left( s\right) \alpha } &=&\stackrel{K-1}{\stackunder{d=0}{\sum }}%
\left[ \frac{\partial ^2S}{\partial q_{\left( u\right) c}\partial q_{\left(
d\right) a}}\left( dq_{\left( d\right) a}-\stackrel{K-1}{\stackunder{s=0}{%
\sum }}\frac{\partial H_{\left( s\right) \alpha }^{\prime }}{\partial
p_{\left( d\right) a}}dt_{\left( s\right) \alpha }\right) +\right. \\
&&\left. +\frac{\partial ^2S}{\partial q_{\left( u\right) c}\partial
t_{\left( d\right) \beta }}\left( dt_{\left( d\right) \beta }-\stackrel{K-1}{%
\stackunder{s=0}{\sum }}\frac{\partial H_{\left( s\right) \alpha }^{\prime }%
}{\partial P_{\left( d\right) \beta }}dt_{\left( s\right) \alpha }\right)
\right] .
\end{eqnarray*}

If the total differential equation given by (\ref{Eq.50}) are valid, the
equation above becomes 
\begin{equation}
dp_{\left( u\right) c}=-\stackrel{K-1}{\stackunder{s=0}{\sum }}\frac{%
\partial H_{\left( s\right) \alpha }^{\prime }}{\partial q_{\left( u\right)
c}}dt_{\left( s\right) \alpha }  \label{Eq.72}
\end{equation}
were, as before, $u=0,1,...,K-1$; $c=0,1,...,N$ and $\alpha =0,1,...,R$.

Making 
$Z\equiv S\left( t_{\left( s\right) \alpha };q_{\left( s\right) a}\right) $
and using the momenta definitions together with equation (\ref{Eq.50}) we
have 
\[
dZ=\stackrel{K-1}{\stackunder{d=0}{\sum }}\frac{\partial S}{\partial
t_{\left( d\right) \beta }}dt_{\left( d\right) \beta }+\stackrel{K-1}{%
\stackunder{d=0}{\sum }}\frac{\partial S}{\partial q_{\left( d\right) a}}%
dq_{\left( d\right) a}, 
\]

\[
dZ=-\stackrel{K-1}{\stackunder{d=0}{\sum }}H_{\left( d\right) \beta
}dt_{\left( d\right) \beta }+\stackrel{K-1}{\stackunder{d=0}{\sum }}%
p_{\left( d\right) a}\left( \stackrel{K-1}{\stackunder{s=0}{\sum }}\frac{%
\partial H_{\left( s\right) \alpha }^{\prime }}{\partial p_{\left( d\right)
a}}dt_{\left( s\right) \alpha }\right) . 
\]

With a little change of indexes we get 
\begin{eqnarray}
dZ=\stackrel{K-1}{\stackunder{d=0}{\sum }}\left( -H_{\left( d\right) \beta }+%
\stackrel{K-1}{\stackunder{s=0}{\sum }}p_{\left( s\right) a}\frac{\partial
H_{\left( d\right) \beta }^{\prime }}{\partial p_{\left( s\right) a}}\right)
dt_{\left( d\right) \beta }.  \label{Eq.63}
\end{eqnarray}

This equation together with equations (\ref{Eq.50}) and (\ref{Eq.72}) are
the total differential equations for the characteristics curves of the HJPDE
given by equation (\ref{Eq.41}) and, if they form a completely integrable
set, their simultaneous solutions determine $S\left( t_{\left( s\right)
\alpha };q_{\left( s\right) a}\right) $ uniquely from the initial
conditions. Besides that, equations (\ref{Eq.50}) and (\ref{Eq.72}) are the
equations of motion of the system written as total differential equations.

\section{Conclusions}

We have obtained the equations of motion for the canonical variables of a
singular higher order system as total differential equations. Each {\it %
coordinate} $q_{\left( s\right) \alpha }\equiv t_{\left( s\right) \alpha }$ $%
\left( \alpha =1,...,R\right) $ is treated as a parameter that describes the
system evolution. The {\it Hamiltonians} $H_{\left( s\right) \alpha
}^{\prime }$ will be the generators of the canonical transformations
parametrized by $t_{\left( s\right) \alpha }$ in the same way the
Hamiltonian $H_0$ is the generator of time evolution. If we have $K=1$ the
results obtained here will reduce to the case of first order systems showed
in ref. \cite{Guler 1}. For $K=2$ we have the same results obtained for a
second order system of ref. \cite{Eu}, where the Hamilton-Jacobi formalism
was applied to Podolsky generalized electrodynamics and the results were
compared to Dirac's Hamiltonian formalism.

The integrability conditions that have to be satisfied by equations (\ref
{Eq.50}), (\ref{Eq.72}) and (\ref{Eq.63}) are analogous to those that have
to be satisfied in the first order case. These conditions have been derived
in ref.\cite{Guler 2} and can be easily applied to the higher order case
developed here. These integrability conditions are equivalent to the
consistence conditions in Dirac's formalism.

We must point out that one of the reasons to consider the constraints'
structure described in Sect. 2 is the fact that if we had considered only
the constraints containing the momenta 
$p_{\left( K-1\right) \alpha }$, given by equations of the form (\ref{Eq19}%
), when developing the singular case in Sect. 4, the {\it coordinate} $%
t_{\left( K-1\right) \alpha }\equiv q_{\left( K-1\right) \alpha }$ would be
an arbitrary parameter in the formalism but the {\it coordinate} $t_{\left(
K-2\right) \alpha }\equiv q_{\left( K-2\right) \alpha }$ (that obeys $%
q_{\left( K-1\right) \alpha }\equiv \stackrel{.}{q}_{\left( K-2\right)
\alpha }$) would have a dynamics of its own. So, when we choose to deal with
all constraints given by expression (\ref{Eq24}) we are avoiding such
contradictions. Furthermore, if we had not made this choice, we would have
an unnecessary extra work when analyzing the integrability conditions since
the constraints involving momenta $p_{\left( s\right) \alpha }$ with $s<K-1$
would appear in this stage imposing extra integrability conditions.

As we mentioned in Introduction, Hamilton-Jacobi formalism is not well
studied for singular systems yet. We still lack a complete analysis of the
relation between the procedures in this new formalism for singular systems
and traditional ones, specially the relation with Dirac's Hamiltonian
formalism. Besides, Hamilton-Jacobi formalism shall be applied to various
physical systems so that we can get a better understanding of its potential
to deal with specific problems.

\section{Acknowledgments}

We thank J. L. Tomazelli for useful suggestions and careful reading of the
ma\-nus\-cript.

B. M. P. is partially supported by CNPq and R. G. T. is supported by CAPES.


\begin{thebibliography}{99}
\bibitem{Dirac 1}  P. M. A. Dirac, Can. J. Math. {\bf 2}, 129 (1950).

\bibitem{Dirac 2}  P. M. A. Dirac, Can. J. Math. {\bf 3}, 1 (1951).

\bibitem{Dirac 3}  P. M. A. Dirac, {\it Lectures on Quantum Mechanics}
(Belfer Graduate School of Science, Yeshiva University, New York, N. Y.,
1964).

\bibitem{Regge}  A. Hanson, T. Regge and C. Teitelboim, {\it Constrained
Hamiltonian Systems} (Accademia Nazionale dei Lincei, Roma, 1976).

\bibitem{Sundermeyer}  K. Sundermeyer, {\it Lecture Notes in Physics 169 -
Constrained Dynamics} (Springer-Verlag, 1982).

\bibitem{Gitman}  D. M. Gitman and I. V. Tyutin, {\it Quantization of Fields
with Constraints} (Springer-Verlag, 1990).

\bibitem{Teitelboim}  M. Henneaux and C. Teitelboim, {\it Quantization of
Gauge Systems} (Princeton University Press, 1992).

\bibitem{Lanczos}  C. Lanczos, {\it The Variational Principles of Mechanics}%
, Fourth Edition (University of Toronto Press, 1970).

\bibitem{Guler 1}  Y. G\"{u}ler, Il Nuovo Cimento B {\bf 107}, 1389 (1992).

\bibitem{Guler 2}  Y. G\"{u}ler, Il Nuovo Cimento B {\bf 107}, 1143 (1992).

\bibitem{Guler 3}  E. M. Rabei and Y. G\"{u}ler, Phys. Rev. A {\bf 46}, 3513
(1992).

\bibitem{Guler 4}  Y. G\"{u}ler, Il Nuovo Cimento B {\bf 109}, 341 (1994).

\bibitem{Guler 5}  Y. G\"{u}ler, Il Nuovo Cimento B {\bf 110}, 307 (1995).

\bibitem{Guler 6}  Y. G\"{u}ler, Il Nuovo Cimento B {\bf 111}, 513 (1996).

\bibitem{Podolsky}  B. Podolsky and P. Schwed, Rev. Mod. Phys. {\bf 20}, 40
(1948).

\bibitem{Bollini}  D. G. Barci, C. G. Bollini and M. C. Rocca, Int. J. Mod.
Phys. A {\bf 10}, 1737 (1995).

\bibitem{Nesterenko}  V. V. Nesterenko, J. Phys. A: Math. Gen. {\bf 22},
1673 (1989).

\bibitem{Batlle}  C. Batlle, J. Gomis, J. M. Pons and N. Rom\'{a}n-Roy, J.
Phys. A: Math. Gen. {\bf 21}, 2693 (1988).

\bibitem{Saito}  Y. Saito, R. Sugano, T. Ohta and T. Kimura, J. Math. Phys 
{\bf 30}, 1122 (1989).

\bibitem{Eu}  B. M. Pimentel and R. G. Teixeira, Il Nuovo Cimento B, {\bf 111%
}, 841 (1996).

\bibitem{Ostrogradski}  M. Ostrogradski, Mem. Ac. St. Petersbourg {\bf 1},
385 (1850).

\bibitem{Mukunda}  E. C. G. Sudarshan and N. Mukunda, {\it Classical
Dynamics: A Modern Perspective} (John Wiley \& Sons Inc., New York, 1974).

\bibitem{Caratheodory}  C. Carath\'{e}odory, {\it Calculus of Variations and
Partial Differential Equations of the First Order}, Part II, page 205
(Holden-Day, 1967).
\end{thebibliography}
\end{document}